\documentclass[final]{elsarticle}

\usepackage{hyperref}

\journal{Journal of \LaTeX\ Templates}

\begin{document}

\begin{frontmatter}

\title{Memory effects and KWW relaxation of the interacting magnetic nano-particles\tnoteref{mytitlenote}}

\author{Ekrem Aydiner} 
\address{Department of Physics, Faculty of Science, \.{I}stanbul University, 34134, \.{I}stanbul, Turkey}


\cortext[mycorrespondingauthor]{Corresponding author}
\ead{ekrem.aydiner@istanbul.edu.tr}


\begin{abstract}
The nano-particle systems under theoretically and experimentally investigation because of the potential applications in the nano-technology such as drug delivery, ferrofluids mechanics, magnetic data storage, sensors, magnetic resonance imaging, and cancer therapy. Recently, it is reported that interacting nano-particles behave as spin-glasses and experimentally show that the relaxation of these systems obeys stretched exponential i.e., KWW relaxation. Therefore, in this study, considering the interacting nano-particle systems we model the relaxation and investigate frequency and temperature behaviour depends on slow relaxation by using a simple operator formalism. We show that relaxation deviates from Debye and obeys to KWW in the presence of the memory effects in the system. Furthermore, we obtain the frequency and temperature behaviour depend on KWW relaxation. We conclude that the obtained results are consistent with experimental results and the simple model, presented here, is very useful and pedagogical to discuss the slow relaxation of the interacting nano-particles. 
\end{abstract}


\end{frontmatter}


\section{Introduction}

The relaxation of the interacting magnetic nano-particle systems are under intensive investigation due to their implications in various fields such as magnetic data storage, cancer therapy, drug delivery, ferrofluids mechanics, magnetic resonance imaging and sensors \cite{Vuong2017,Basel2012,Huang2013,Ayala2013,Kenouche2014,Amos2012,Teresa2014,Estelrich2015,Sousa2017,Ortega2013,Zhao2012}. It is known that the ideal non-interacting magnetic nano-particles are composed of a single magnetic domain which has super-paramagnetic properties. Their diameter is below 3–50 nm, depending on the materials and the relaxation of the interacting magnetic nano-particle is generally given by Debye exponential \cite{Debye1929}. However, it was shown in a many experimental studies such as Co$_x$Ag$_{1-x}$  nano-particles \cite{Zhang2007}, $\gamma$-Fe$_{2}$O$_{3}$ nano-particles \cite{Gazeau1998}, Fe$_{3}$N nano-particles \cite{Mamiya1998} that the relaxation deviates from Debye type exponential relaxation and obey to $\sim\exp(-t/\tau)^{\alpha}$ where $\alpha$ is the decaying parameter. This slow stretched exponential relaxation is often called as, historically, Kohlrausch-Williams-Watts (KWW) law \cite{Kohlrausch1847,Williams1970}. Several theoretical mechanism have been proposed to explain non-exponential KWW relaxation behavior of the nano-particles \cite{Zhang2007,Gazeau1998,Mamiya1998,Parker2005,Bean1963,Wohlfarth1980,Street1959,Dattagupta1987,Dattagupta2014}. In these theories, mainly, the concentrated nano-particle systems behave as spin glasses \cite{Cannella1972,Kirkpatrick1978,Jonscher1977} due to the dipole-dipole interaction of nano-particles \cite{Parker2005}, the cooperative behavior \cite{Mamiya1998} and the long-range interactions. The memory effects arise depends on these puzzling dynamics and which leads to KWW relaxation. 

The KWW relaxation is characteristically different from the exponential type, which can lead to amazing physical behaviours. The effect of this difference in relaxation shows itself in the response function. For instance, the frequency and temperature behaviour of the reactive and the dissipation part of the response function can dramatically change. This effect may be too significant to be neglected for some applications in the nano-scale. Therefore, modelling of the KWW relaxation of the nano-particles and showing this difference can be a guide for applications in practice. Based on this motivation, in this study, we focus on the modelling of the relaxation for the interacting nano-particle system. By using a simple operator formalism we will
discuss relaxation function and the frequency and temperature-dependent behaviour of the interacting nano-particles. 

In Section 2, we present the relaxation of the non-interacting nano-particle based on a two-level jumping model by using operator formalism. In Section 3, we discuss the relaxation function of the interacting nano-particle system with a multi-levels jumping process.
In the same section, we obtain correlation and relaxation function for interacting nano-particle systems and we discussed frequency and temperature dependence of the model system. The last Section is devoted the conclusion. 

\section{The non-interacting magnetic nano-particles}

In this section, we briefly review the simple model for the paramagnetic relaxation under an external field \cite{Neel1950,Brown1963} and the simple operator formalism with two-level jumping \cite{Dattagupta1987}. 
The magnetic energy of the particle has
its origin in the anisotropy energy that is associated with the
crystalline structure of the material. For uniaxial particles, the
magnetic anisotropy energy $E_{a}$ may be written as
\begin{equation} \label{n-int-Ea}
	E_{a}=KV\left( 1-n_{z}^{2}\right)
\end{equation}
where $n_{z}$ is the component of $\widehat{n}$ along the $z$-axis,
which is chosen to label the direction of anisotropy, $K$ is a
constant that depends on material properties, and $V$ is the volume
of the particle. Denoting by $\theta$ the angle between $z$ and
$\widehat{n}$, anisotropy energy $E_{a}$ can be written as
\begin{equation} \label{n-int-Etheta}
	E(\theta)=KV\sin^2\theta \ .
\end{equation}
However, when the particle is exposed to the external field, the
energy contribution to the anisotropy energy comes from external
field $H$. This energy is known Zeeman energy which is given by
\begin{equation} \label{n-int-E-Z}
	E_{Z}=VM_{0} H \widehat{n} .
\end{equation}
Therefore, total energy of non-interacting superparamagnetic
particle is given by
\begin{equation} \label{n-int-E-gt}
	E(\theta)=KV\sin^2\theta-VM_{0}H_{0} \cos\theta.
\end{equation}
The orientation probability of the magnetic nano-particle under
external field depends on energy which is given by
\begin{equation} \label{n-int-prob}
	p\left( \theta\right)  =\frac{\exp\left[  -\beta E\left(
		\theta\right) \right]}{\sum_{\theta}\exp\left[  -\beta E\left(
		\theta\right) \right]} .
\end{equation}
But, the symmetry is broken by an external field $H_{0}$ along the
$z$ axis, hence, it is supposed that the orientation $\theta=0$ is
more probable, now that the particle finds it energetically more
favorable to line up along the magnetic field. The minima of the
energy occur at $\theta=0$ and around $\theta=\pi$, which define the
two equilibrium orientations of the particle. 

In this case, it is
assumed that the particle is under the constant influence of
spontaneous thermal fluctuations. Once in a while, these
fluctuations (or "kicks" from thermal phonon, loosely speaking) are
strong enough to enable the particle to overcome the barrier between
$\theta=0$ and around $\theta=\pi/2$. Essentially, the particle has
to cross a potential hump whose height is $E_{min}-E_{max}=KV$. The
particle remains in one of its equilibrium positions most of the
time; occasionally it undergoes an instantaneous jump from one
equilibrium orientation to another. This two-level stochastic process is so-called Kubo-Anderson process that
mathematically expressed by the equilibrium probabilities clear from
Eq.\,(\ref{n-int-prob})
\begin{equation} \label{n-int-prob-12}
	p_{1,2}=p\left(  \theta=0, \pi \right)  =\frac{\exp\left[  -\beta
		E_{1,2}\right] }{\exp\left[  -\beta E_{1}\right]  +\exp\left[  -\beta
		E_{2}\right]  }
\end{equation}
for two-level jumping process. On the other hand, energies $E_{1}$
and $E_{2}$ are
\begin{equation} \label{n-int-prob-E12}
	E_{1,2}=E\left( \theta =0,\pi \right) = \mp VM_{0}H_{0}
\end{equation}
Using single activation energy arguments, the transition rates per
unit time (due to thermal fluctuations) may be written as
\begin{equation} \label{n-int-prob-W0}
	\widehat{W}_{12,21}=\nu _{0}\exp \left[ -\beta \left( E_{max}-E_{2,1}\right) %
	\right]
\end{equation}
where $E_{max}$ is the maximum value of the energy at the hump, and
$\nu _{0}$ is the attempt frequency. We assume here of course that
the attempt frequency $\nu _{0}$ stays the same even in the presence
of $H_{0}$. It is clear from Eq.\,(\ref{n-int-E-gt}) that the maximum in energy
occurs at
\begin{equation} \label{n-int-theta}
	\cos\theta_{max}=-M_{0}H_{0}/2K
\end{equation}
and therefore, the maximum energy is defined as
\begin{equation} \label{n-int-emax}
	E_{max}=KV\left[1+\left(\frac{M_{0}H_{0}}{2K}\right)^2\right] \ .
\end{equation}
This relaxation process can be modelled by using the operator formalism with two level jumping \cite{Dattagupta1987}. In the operator formalism, states $\left\vert n \right\rangle$ for two-level system are given in the form
\begin{equation} \label{n-int-states}
	\left\vert 1\right\rangle =\left(
	\begin{array}
		[c]{c}%
		1\\
		0
	\end{array}
	\right), \ \left\vert 2\right\rangle =\left(
	\begin{array}
		[c]{c}%
		0\\
		1
	\end{array}
	\right) . 
\end{equation}
where $|...\rangle$ represents a stochastic state like the Dirac ket. Jumping of the stochastic variable $\mu$ between two values are given in the tensor form as
\begin{equation} \label{n-int-epsilon}
	\widehat{\mu}= \left(
	\begin{array}
		[c]{cc}%
		\mu_{1} & 0 \\
		0 & \mu_{2} 
	\end{array}
	\right) .
\end{equation}
In the absence of external magnetic field, the two directions have completely equivalent probability. In this case process completely reduce to simple two-levels jumping process. In any case, the jumping between these states is governed by the Markovian master equation  \cite{Dattagupta1987}
\begin{equation} \label{master1}
	\frac{\partial}{\partial t} \widehat{P}(t)  = \widehat{P}(t)\widehat{W}
\end{equation}
where $\widehat{P} (t)$ is the transition probability of the $i$th site and $\widehat{W}$ is the transition rate which is given by
\begin{equation}\label{n-int-W01}
	\widehat{W}_{12}=\lambda p_{1}, \quad \widehat{W}_{21}=\lambda p_{2}
\end{equation}
where $\lambda$ is the relaxation rate of the system, which can be
represented in terms of $E_{max}$
\begin{equation} \label{n-int-lambda}
	\lambda=\lambda_{0} \exp \left[-\beta E_{max}\right]
\end{equation}
where $\lambda_{0}$ is a constant. The rate at which such
jumps occur is related to difference between the maximum and minimum
values of energy in the system. 

The transition probabilities in
$\widehat{W}_{21} = \lambda p_{2}$ and $\widehat{W}_{12} = \lambda
p_{1}$ are consistent with the detailed balance relation which reads
\begin{equation} \label{n-int-detailed}
	p_{1} \widehat{W}_{21}=p_{2} \widehat{W}_{12} \ .
\end{equation}
The transition rate $\widehat{W}$ is given in terms of collision matrix $\widehat{T}$ as
\begin{equation} \label{n-int-transition}
	\widehat{W}  = \lambda\left(  \widehat{T}-\mathbf{1}\right),
\end{equation}
where $\lambda = 2w$ is the eigenvalue, $\widehat{T}$ is the collision matrix 
and $\mathbf{1}$ is the unit matrix \cite{Dattagupta1987}. Here, $\widehat{T}$ is specified to be
\begin{equation} \label{n-int-lambda-J}
	\widehat{T}  = \left(
	\begin{array}
		[c]{ccc}%
		p_{1} & p_{1}  \\
		p_{2} & p_{2} 
	\end{array}
	\right),
\end{equation}
where $p_{m}$ corresponds to transition probability between states or levels ($m=1,2$) and satisfy $p_{1}+p_{2}=1$.
The new process is defined by the relation
\begin{equation} \label{n-int-j-expect}
	p_{m} = \left\langle m\right\vert \widehat{T}\left\vert n\right\rangle , \quad n,m=1,2 \ .
\end{equation}
For the two-level relaxation process, $\widehat{T}$ is idempotent. This is so because
\begin{equation} \label{n-int-j-square}
	p_{m} = \left\langle m\right\vert \widehat{T}^{2}\left\vert n\right\rangle
	=\left\langle
	m\right\vert \widehat{T}\left\vert n\right\rangle.
\end{equation}
By using this property, the conditional probability matrix $\widehat{P}(t)$ can be constructed. For Markovian processes, conditional probability is simply obtained from Eq.\,(\ref{master1}) together with Eqs.\,(\ref{n-int-lambda-J})-(\ref{n-int-j-square}) in the form
\begin{equation} \label{n-int-conP}
	\widehat{P}(t) = \exp(\widehat{W }t) = \exp(-\lambda t) \left[ \mathbf{1} - \widehat{T} + \widehat{T} \exp (\lambda t) \right] . 
\end{equation}
By using above formalism, the correlation function can ne carried out as
\begin{equation} \label{n-int-Corr}
	C(t) = \int \int d\mu d\mu_{0} \mu_{0} p(\mu_{0}) \mu \langle\mu|\widehat{P}(t)|\mu_{0} \rangle .
\end{equation}
For two-level Kubo-Anderson process this correlation function is governed by the Néel relaxation \cite{Neel1950}. Many physical properties of the non-interacting nano particles can be computed by this correlation function.

\section{The interacting magnetic nano-particles}

In the case of the interacting magnetic nano-particles, the relaxation deviates from Debye type relaxation and obeys KWW type relaxation depends on the presence of the meta-stable states due to the magnetic viscosity, the long-range interactions which can lead to the glassy behaviour. Therefore, the relaxation can be modelled by multi-levels jumping instead of the two-levels Kubo-Anderson process. New relaxation picture to the modelling of interacting nano-particles sometimes is called as s kangaroo process \cite{Dattagupta2014} which serves a simple formalism the relaxation of the interacting magnetic nano-particles. Below, we briefly give the formalism and numerical results of the model.

\subsection{Generalized formalism}

The meta-stable state energy levels in the energy phase spaces of the interacting nanoparticles probably lead to the discrete jumping process. In this case, the relaxation can be modelled as the multi-level jumping process where $N\rightarrow \infty$. Thus, the process may be regarded as continuous. The jump matrix $\widehat{W}$ for the multi-level jumping process is given by
\begin{equation} \label{int-W-jump}
	\langle \mu|\widehat{W}|\mu_{0} \rangle = \lambda \left[p(\mu) - \delta(\mu-\mu_{0})  \right] 
\end{equation}
where the jump rate $\lambda$ corresponds to the constant rate. However, in the presence of the memory effects, Eq.\,(\ref{int-W-jump}) must be extended to
\begin{equation} \label{int-W-exp}
	\langle \mu|\widehat{W}|\mu_{0} \rangle = \lambda(\mu_{0}) \left[q(\mu) - \delta(\mu-\mu_{0})  \right]
\end{equation}
where $q(\mu)$ does not take constant values for each jumping, however, it satisfy
\begin{equation} \label{int-sum-q}
	\int d\mu  q(\mu) = 1 \ .
\end{equation}
For present case, the detailed balance relation in Eq.\,(\ref{n-int-detailed}) is given by 
\begin{equation} \label{int-det-balance}
	p(\mu_{0}) \lambda(\mu_{0}) q(\mu) = p(\mu) \lambda(\mu)  q(\mu_{0}) \ .
\end{equation}
The summation over the $\mu_{0}$ gives 
\begin{equation} \label{int-q}
	q(\mu) = \frac{p(\mu) \lambda(\mu) }{\int d\mu_{0} p(\mu_{0}) \lambda(\mu_{0}) } \ .
\end{equation}
The jump matrix is given as
\begin{equation} \label{int-W}
	\widehat{W} = \widehat{\Lambda} \left(\widehat{T}-\mathbf{1}\right)
\end{equation}
where the matrix of $\widehat{T}$ is 
\begin{equation} \label{int-exp-J}
	\langle \mu|\widehat{T}|\mu_{0} \rangle =  q(\mu)
\end{equation}
and $\widehat{\Lambda}$ is diagonal and is given by
\begin{equation} \label{int-exp-Lam}
	\langle \mu|\widehat{\Lambda}|\mu_{0} \rangle =  \lambda(\mu_{0}) \delta(\mu- \mu_{0}) \ .
\end{equation}
Evidently, Eqs.\,(\ref{int-W}) and (\ref{int-exp-Lam}) are consistent with results of Kubo-Andreson  model. The question now is how to obtain $\widehat{P}(t)$. Here we have to note that the memory effects can be considered in the continuous-time random walk (CTRW) framework \cite{Montroll1965,Scher1973}. Discrete dynamics are categorized by the probability density function (pdf). 
In the more general case, any finite characteristic waiting time is given by
$T=\int_{0}^{\infty} t \psi(t) dt$ and any finite jump length variance is given by $\Sigma^{2} =\int_{-\infty}^{\infty} x^{2} \lambda(x) dx$. The corresponding process in the diffusion limit shows normal diffusive behaviour with Gaussian pdf \cite{Montroll1965,Scher1973,Metzler2000,Metzler2002}.  
In a simple random walk process, the waiting time pdf
$\psi(t)$ is of Poisson form and jump length pdf $\lambda(x)$ is of Gaussian form. 
However, the waiting time $T$ diverges, conversely, the jump length variance $\Sigma^{2}$ is still kept finite for the non-Markovian process. In a such process, long-tailed waiting time pdf takes an asymptotic form \cite{Metzler2000}. Thus, the CTRW process with power-law form $\psi(t) \approx t^{-1-\alpha}$ where $0<\alpha<1$, leads to the fractional diffusion equation in the continuum limit \cite{Montroll1965,Scher1973,Metzler2000,Metzler2002}. By using this approximation Eq.\,(\ref{n-int-conP})
can be obtained. However, here, we will use simple operator formalism presented above instead of the fractional diffusion approach. 

\subsection{Correlation function}

To obtain $\widehat{P}(t)$ we follow a simple procedure given in Ref.\,\cite{Dattagupta2014}. By using Eqs.\,(\ref{int-W}) and  (\ref{int-exp-Lam}) Laplace transformation is given by
\begin{equation} \label{int-P-Laplace}
	\tilde{P}(s) = \int_{-\infty}^{\infty} e^{st} \widehat{P}(t) dt
\end{equation}
where $s$ is a complex number frequency parameter, which leads to 
\begin{equation} \label{int-P-Laplace-2}
	\tilde{P}(s) = \frac{1}{ (s + \widehat{\Lambda}) - \widehat{T} \widehat{\Lambda}} \ .
\end{equation}
The Eq.(\ref{int-P-Laplace-2}) is written as a geometric series
\begin{eqnarray} \label{int-P-geo}
	\tilde{P}(s) = \frac{1}{ s + \widehat{\Lambda} } \left[ 1 + (\widehat{T}\widehat{\Lambda}) \frac{1}{(s + \widehat{\Lambda})} +  (\widehat{T}\widehat{\Lambda}) \frac{1}{ (s + \widehat{\Lambda})}(\widehat{T}\widehat{\Lambda}) \frac{1}{ (s + \widehat{\Lambda})}+ ...  \right] \ .
\end{eqnarray}
Hence, the expectation value of the $\tilde{P}(s)$ is given by
\begin{eqnarray} \label{int-exp-Ps}
	\langle \mu|\tilde{P}(s)|\mu_{0} \rangle = \frac{1}{ s + \lambda(\mu_{0}) }  \left[ \delta(\mu-\mu_{0}) + (\mu|\widehat{T}\widehat{\Lambda}|\mu_{0}) \frac{1}{ s + \lambda(\mu)} +  ...  \right] \ .
\end{eqnarray}
Here, we have used the completeness relation
\begin{equation} \label{int-closeure}
	\int d\mu |\mu \rangle \langle \mu| = 1 \ .
\end{equation}
We note that 
\begin{equation} \label{int-exp-TL}
	\langle \mu|\widehat{T}\widehat{\Lambda}|\mu_{0} \rangle = \int d\mu^{\prime} q(\mu) \lambda(\mu_{0}) \delta(\mu_{0}-\mu^{\prime}) = \lambda (\mu_{0}) q(\mu)  \ .
\end{equation}
Hence, Eqs.\,(\ref{int-exp-Ps}) yields
\begin{equation} \label{int-exp-Ps-sum}
	\langle \mu|\tilde{P}(s)|\mu_{0} \rangle =\frac{1}{ s + \lambda(\mu_{0}) } \left[ \delta(\mu-\mu_{0}) + \lambda(\mu_{0}) \frac{q(\mu)}{s + \lambda(\mu)} +  ...  \right]
\end{equation}
The higher order terms can be dropped and $q(\mu)$ becomes $p(\mu)$, hence, above equation reduce to 
\begin{equation} \label{int-exp-Ps-reduce}
	\langle \mu|\tilde{P}(s)|\mu_{0} \rangle = \frac{1}{ s + \lambda } \left[ \delta(\mu-\mu_{0}) +  \frac{\lambda}{s} p(\mu)  \right]
\end{equation}
which yields the Laplace transform of the probability matrix for
the Kubo–Anderson process. By using Eq.\,(\ref{int-exp-Ps-reduce}), the correlation function in Eq.\,(\ref{n-int-Corr})  is given by 
\begin{equation} \label{int-Corr-Laplace}
	\tilde{C}(s) = \int d\mu  \frac{p(\mu) \mu^{2}}{ s + \lambda (\mu)} + ... \ .
\end{equation}
Transforming back to the time domain, the correlation function can be written as
\begin{equation} \label{int-Corr-time}
	C(t) =   \mathcal{L}^{-1} \{ \tilde{C}(s) \}  = \frac{1}{2\pi i} \int \tilde{C}(s) e^{-\lambda(\mu) s t} ds
\end{equation}
which is now a continuous superposition of exponentially decaying functions of time. 
Finally, we obtain
\begin{equation} \label{int-Corr-fin}
	C(t) = \left\langle \mu^{2} \right\rangle \exp \left[ - \left(\left\langle\lambda(\mu)\right\rangle  t \right)^{\alpha}  \right] = \frac{V^{2} M_{0}}{k_{B}T}  \exp \left[ - \left( \lambda_{eq}  t\right)^{\alpha}  \right],  \quad 0<\alpha < 1
\end{equation}
where $\lambda_{eq}$ can be chosen as $\lambda$. This choice does not change of the nature of the problem since relaxation process is dominated by $E_{max}$ in the system.

\subsection{The KWW relaxation}

\begin{figure} [ht!]
	\centering
	\includegraphics[width=12cm]{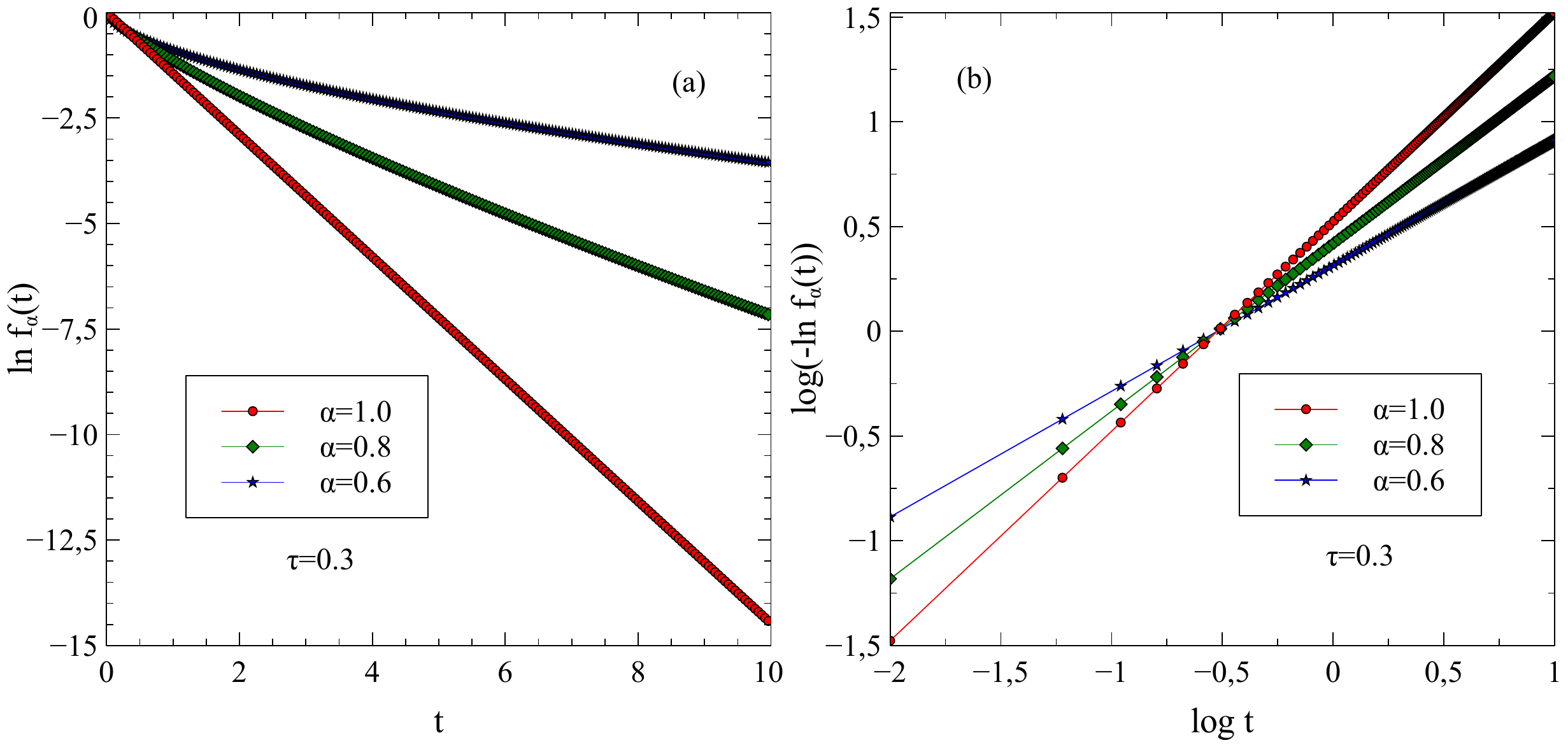}
	\caption{The KWW relaxation function $f_{\alpha}(t)$ with respect to time for different $\alpha$ values where $f_{0}=\frac{V^{2} M_{0}}{k_{B}T} =1$.  In (a) the relaxation function $f_{\alpha}(t)$ in ln-linear scale is shown. In (b), the relaxation function $f_{\alpha}(t)$ in log(-ln)-log scale is provided.}
	\label{fig-relax}
\end{figure}

The KWW relaxation function is easily obtained from the correlation function Eq.\,(\ref{int-Corr-fin}) as
\begin{equation} \label{int-kww}
	f_{\alpha}(t) =  \frac{V^{2} M_{0}}{k_{B}T}  \exp \left[ - \left(\frac{t}{\tau}\right)^{\alpha}  \right],  \quad 0<\alpha < 1
\end{equation}
where $\tau=1/\lambda$. This relaxation is known as KWW relaxation or stretched exponential \cite{Kohlrausch1847,Williams1970}. Eq.\,(\ref{int-kww}) reduces to exponential relaxation of Debye \cite{Debye1929}. For various $\alpha$, the relaxation curves of the interacting nano-particles at fixed $\tau=0.3$ is given in Fig.\ref{fig-relax}. The exponential relaxation for $\alpha=1$ is denoted with the red circle Fig.\ref{fig-relax}(a). However, as can be seen from Fig.\ref{fig-relax}(a) that the curves characteristically deviate from exponential for $\alpha < 1$ (see green and blue lines). One can see that these curves in the log(-ln)-log scale in Fig.\ref{fig-relax}(b) clearly reflects KWW relaxation for $\alpha < 1$.

As we mentioned above that $\alpha$ represents the memorial effects which appear due to the dipole-dipole interaction of nano-particles \cite{Parker2005}, the cooperative behaviour \cite{Mamiya1998} and the long-range interactions. We show that the non-exponential relaxation of the interacting nano-particles by using the presented simple model.

\subsection{The Cole-Cole behaviour}

Now, we discuss the magnetic susceptibility of the magnetic nano-particle systems depends on the KWW relaxation. 
The frequency-dependent susceptibility function can be obtained from integral transformation of the Eq.~(\ref{int-kww}) as
\begin{equation} \label{n-relax-Jw}
	\chi_{\alpha}(w) =  \int_{0}^{\infty} \exp(-iwt) \left[ -\frac{d f_{\alpha} (t)}{dt} \right] dt \ .
\end{equation}
Eq.~(\ref{n-relax-Jw}) can decompose to
\begin{equation} \label{i-relax-J}
	\chi_{\alpha}(w) = \chi^{\prime}_{\alpha}(w) +  i \chi^{\prime\prime}_{\alpha} (w) 
\end{equation}
where the real term corresponds to the reflected or emitted part of the applied external field and the imaginary part denotes the absorbed part by the system. When $\alpha < 1$, there is no the analytical expression of the Eq.\,(\ref{n-relax-Jw}) for real and imaginary parts \cite{Montroll1984,Alvarez1993,Alvarez1991}. The best way is the numerically perform the Fourier transform for the integral except the points $\alpha \ne 1$ and $\alpha\ne 0.5$. Another approximation is the Laplace transformation of the Eq.\,(\ref{n-relax-Jw}). The magnetic susceptibility can be given in terms of the Laplace transformation form \cite{Metzler2002} as
\begin{equation} \label{Ralf-Laplace}
	\chi_{\alpha}(w) = \left[ 1- s \tilde{f}(s) \right]_{s=iw}
\end{equation}
where $\tilde{f}(s)$ is the Laplace transformation which is given as $\tilde{f}(s)=\int_{0}^{\infty} (-d f(t)/dt) e^{st}dt$ followed by the rotation $s\rightarrow iw$ \cite{Metzler2002}.

Following Eq.\,(\ref{Ralf-Laplace}), Laplace transformation can be given in the form 
\begin{equation} \label{n-CC}
	\chi_{\alpha}(w)  =  \chi_{0}  \frac{1}{1 + (iw \tau )^{\alpha} }, \quad 0<\alpha < 1,
\end{equation}
where $\chi_{0} = \frac{V^{2} M_{0}}{k_{B}T}$. This solution is known as the Cole-Cole pattern \cite{Cole1941}. The interval $0<\alpha\le 1$ has been widely used to modify Eq.\,(\ref{i-relax-J}) in order to phenomenologically fit experimental data for the complex compliance. However, in the limits of $\alpha=1$ both the
KWW form Eq.\,(\ref{int-kww}) and  (\ref{n-CC}) reduce to the corresponding classical result which corresponds to Debye type relaxation and
Eq.\,(\ref{i-relax-J}). It is report in Ref.\,\cite{Metzler2002} that a dynamic framework which leads to relaxation functions of the Mittag-Leffler type \cite{Metzler2000,Metzler2002,Weron1996,Hilfer2002,Hilfer2002b,Kalmykov2004,Garrappa2016,Riesz1949,Caputo1967}. 

The complex compliance for KWW relaxation  corresponding to the Mittag-Leffler pattern, the result being exactly the Cole-Cole function (\ref{n-CC}), as obtained earlier by Weron and Kotulski \cite{Weron1996} in a similar context. After several step, the real part and the imaginary part are given \cite{Metzler2002} as
\begin{equation} \label{J-real-p}
	\chi^{\prime}_{\alpha}(w) = \chi_{0}  \frac{1 + (w\tau)^{\alpha} \cos(\pi \alpha /2) }{1 + (w\tau)^{2\alpha} + 2 (w\tau)^{\alpha} \cos(\pi \alpha/2)  } 
\end{equation}
and
\begin{equation} \label{J-imaginary-p}
	\chi^{\prime\prime}_{\alpha}(w) = \chi_{0}  \frac{(w\tau)^{\alpha} \sin(\pi \alpha /2) }{1 + (w\tau)^{2\alpha} + 2 (w\tau)^{\alpha} \cos(\pi \alpha/2)  } \ .
\end{equation}
\begin{figure} [ht!]
	\centering
	\includegraphics[width=7.5cm]{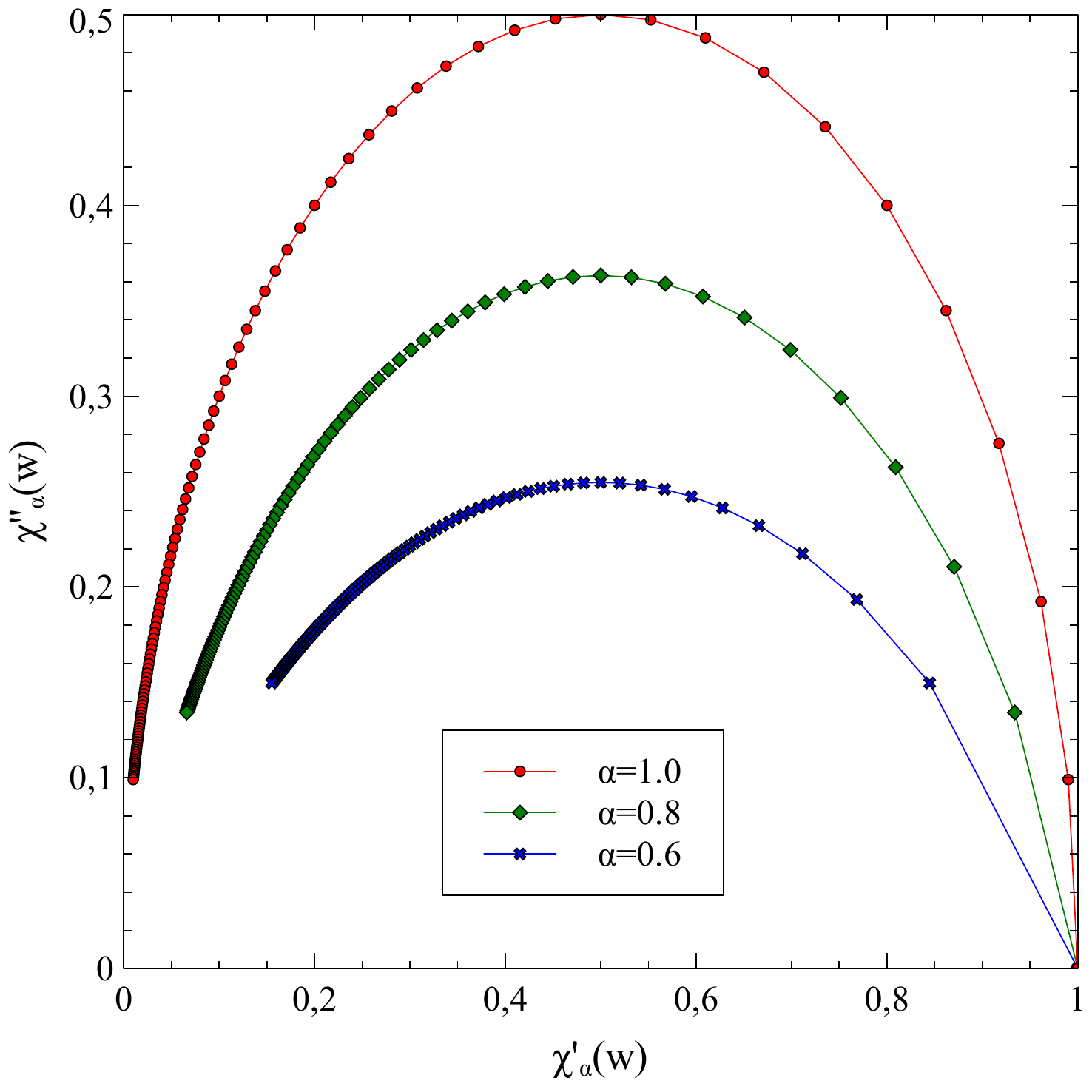}
	\caption{Cole–Cole plot for the magnetic susceptibility of the interacting nano-particles.}
	\label{fig-CC}
\end{figure}
The Cole-Cole plots for the various $\alpha$ values are given in Fig.\ref{fig-CC}. Numerical results in the Fig.\ref{fig-CC} show that the Cole-Cole relation also changes depend on $\alpha$ parameter. This is an expected result, however, which is meaningful since it establishes a link between the interacting nano-particle system and anomalous relaxation.

\subsection{Temperature dependence}

The temperature dependence of the real and imaginary part can be obtained from Eqs.\,(\ref{J-real-p}) and (\ref{J-imaginary-p}) as
\begin{equation} \label{J-real-part}
	\chi^{\prime}_{\alpha}(T)=\chi_{0}  \frac{(w\tau_{0})^{-\alpha} e^{-\alpha x}  + \cos(\pi \alpha/2)}{(w\tau_{0})^{-\alpha} e^{-\alpha x} + (w\tau_{0})^{\alpha} e^{ \alpha x} + 2 \cos(\pi \alpha/2)  } 
\end{equation}
\begin{equation} \label{J-imaginary-temp}
	\chi^{\prime\prime}_{\alpha}(T) = \chi_{0}\frac{\sin(\pi \alpha /2) }{(w\tau_{0})^{-\alpha} e^{-\alpha x} + (w\tau_{0})^{\alpha} e^{ \alpha x} + 2 \cos(\pi \alpha/2)  } 
\end{equation}
where 
\begin{equation} \label{J-x}
	x= \frac{E_{max}}{k_{B}} \frac{T}{T_{g}-T}
\end{equation}
denotes the Vogel-Tammann-Fulcher law \cite{Vogel1921,Tammann1926,Fulcher1925} which fits well to critical slowing down. By using these relations, the magnitude of the susceptibility can be obtained as 
\begin{equation} \label{Q-bekir}
	|\chi_{\alpha}(T) | = \frac{M(T)}{H} =  \left[(\chi^{\prime}_{\alpha}(T))^{2} +  (\chi^{\prime\prime}_{\alpha} (T))^{2} \right]^{1/2}
\end{equation}

\begin{figure} [ht!]
	\centering
	\includegraphics[width=12cm]{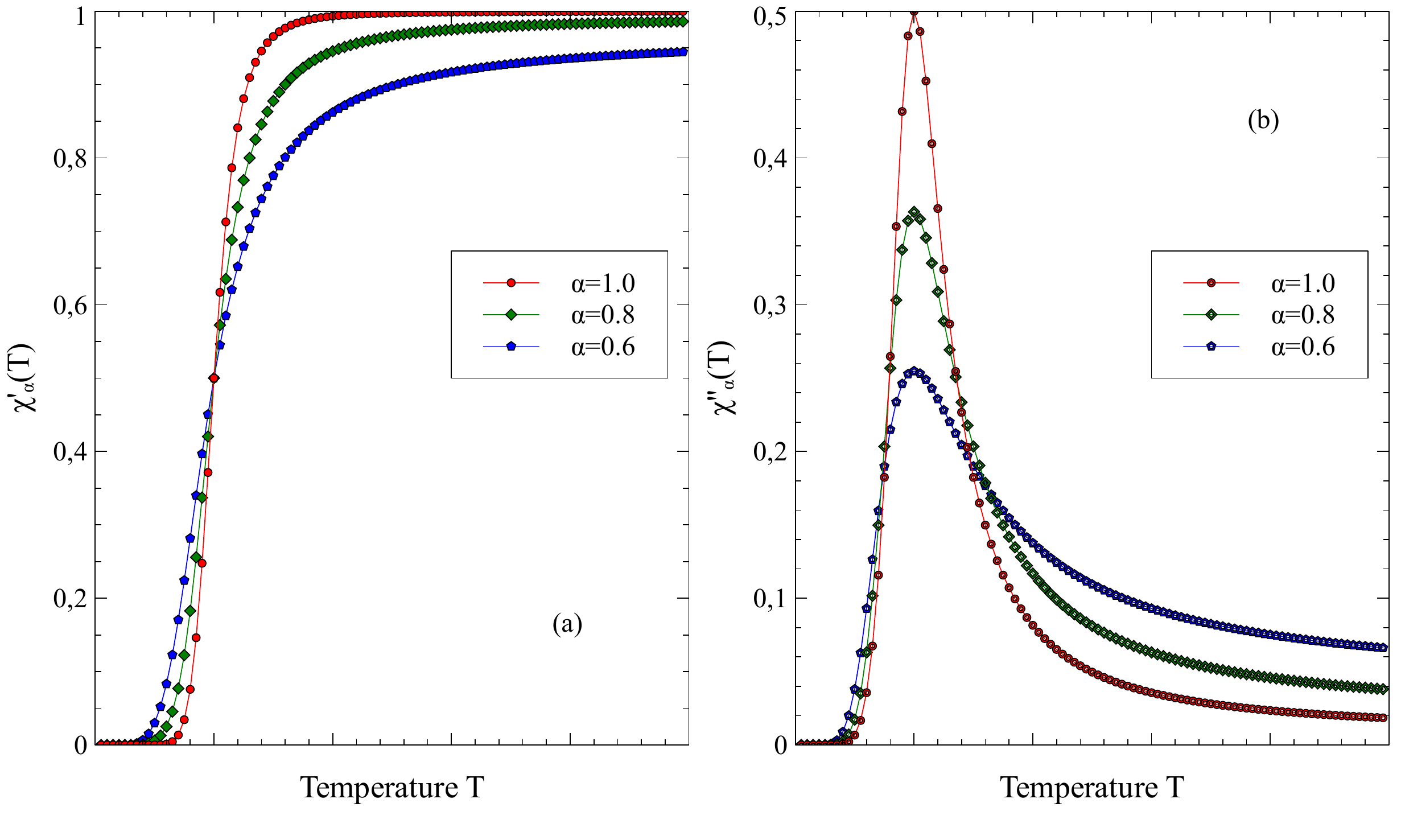}
	\caption{In (a), the real part $\chi^{\prime}(T)$ and in (b) the imaginary part $\chi^{\prime\prime}(T)$ of the magnetic susceptibility for various $\alpha$ values. We set $E_{max}=5.0$, $w=1.0$, $\tau_{0}=1.0$, $T_{g}=1.0$ and Boltzmann constant $k_{B}=1.0$.}
	\label{fig-temp}
\end{figure}

In order to see temperature dependence behaviour of the magnetic susceptibilities, the numerical solutions of Eqs.\,(\ref{J-real-part}) and (\ref{J-imaginary-temp}) for various $\alpha$ values and arbitrary dimensionless parameters $E_{max}=5.0$, $w=1.0$, $\tau_{0}=1.0$, $T_{g}=1.0$ and Boltzmann constant $k_{B}=1.0$ are given in Fig.\,\ref{fig-temp}. 

As can be seen from Fig.\,\ref{fig-temp}(a) real part of the susceptibility rapidly increases with increasing temperature and reach up to the saturation value. Unlike the real part, the imaginary part, in Fig.\,\ref{fig-temp}(b), rapidly increases and reach up a maximum which appears at glassy temperature $T_{g}$, and it decreases non-exponentially. This characteristic KWW type behaviour can be clearly seen in the imaginary part of the susceptibility after, particularly, the glassy temperature $T_{g}$. 
The glassy behaviour of the magnetic nano-particle originates from memory effects due to discrete distribution of relaxation times, coupling and correlations due to weak or strong long-range interactions or topological reasons \cite{Cannella1972,Kirkpatrick1978,Jonscher1977,Richert1994,Phillips1996,Potuzak2011,Grassberger1982,Wu2016}.

Our numerical results for this simple model are consistent with the recent experimental studies. Indeed, similar characteristic behaviour for the real and imaginary part of the susceptibility has been found in the susceptibility measurements of the Fe$_3$O$_4$ nanoparticles and in NiFe$_2$O$_4$ nanoparticles \cite{Maldonado2013,Nadeem2011}.

\section{Conclusion}

In this work, considering the interacting nano-particles we analyse non-exponential KWW relaxation, frequency and temperature dependence of the magnetic susceptibilities of these systems by a simple operator formalism. Firstly, we briefly present the relaxation model for the non-interacting nano-particles based on two-level jumping process. Then, we suggest that the relaxation process of the non-interacting nano-particle systems can be modelled by using multi-level jumping approach since the multi-level discrete energy levels can appear in the system depend on the meta-stable occurs due to the dipole-dipole interaction of nano-particles, the cooperative behaviour and the long-range interactions. We assume that the process may be regarded as continuous in the limit of $N\rightarrow\infty$. By using the generalized method in Ref.\,\cite{Dattagupta2014} we obtain the KWW relaxation function of the interacting nano-particle system. Later, we discuss frequency and temperature dependence of the magnetic susceptibilities depend on KWW exponent $\alpha$ which is a measure of the memory effects in the complex and disordered structure of the non-interacting nano-particles. Obtained numerical results from the simple model presented here are consistent with the experimental results. 

Here, the exponent $\alpha$ can be regarded as a degree of the concentration which can serve as a source of the dipole-dipole interaction of nano-particles, the cooperative behaviour and the long-range interactions in the system. In this scenario, when the concentration is increased in the system, the value of the parameter $\alpha$ decreases. All result shows that the physical behaviour of the interacting nano-particle system completely different from the non-interacting case. 

We again state that the interacting nano-particles are used in many different areas. Therefore, understanding of the spin-glass like behaviour, non-exponential relaxation mechanism and other thermodynamic properties of these systems are very important for many areas such as magnetic data storage, cancer therapy, biomedicine, drug delivery, ferrofluids mechanics, magnetic resonance imaging and sensors. Here, in this study, considering a very simple and pedagogical model we obtain the well-known results for the interacting nano-particles. 

Finally, we can conclude that the very simple theoretical method presented here can be used to the model the physical behaviour of the different interacting systems.

\section*{Acknowledgements}

I would like to thank referee for stimulating guidance which enhanced the quality of the paper.


\end{document}